# EFFECTS OF ULTRASOUND WAVES INTENSITY on the REMOVAL of CONGO RED COLOR from the TEXTILE INDUSTRY WASTEWATER by $Fe_3O_4@TiO_2$ CORE-SHELL NANOSPHERES.

Hossein Ghaforyan[a], Tooraj Ghaffary[*b], Richard Pincak[c], Majid Ebrahimzadeh[a]

[a] Department of Physics, Payame Noor University, I.R. of Iran; Department of Science, Shiraz Branch, Islamic Azad University, Shiraz, Iran. [c] Institute of Experimental Physics, Slovak Academy of Sciences, Watsonova 47,043 53 Kosice, Slovak Republic and Bogoliubov Laboratory of Theoretical Physics, Joint Institute for Nuclear Research, 141980 Dubna, Moscow region, Russia

**Abstract:** Environmental pollutants, such as colors from the textile industry, affect water quality indicators like color, smell, and taste. These substances in the water cause the obstruction of filters and membranes and thereby reduce the efficiency of advanced water treatment processes. In addition, they are harmful to human health because of reaction with disinfectants and production of by-products. Iron oxide nanoparticles are considered effective absorbents for the removal of pollutants from aqueous environments. In order to increase the stability and dispersion, nanospheres with iron oxide core and titanium dioxide coating were used in this research and their ability to absorb Congo red color was evaluated. Iron oxide-titanium oxide nanospheres were prepared based on the coprecipitation method and then their physical properties were determined using a tunneling electron microscope (TEM) and an X-ray diffraction device. Morphological investigation of the absorbent surface showed that iron oxide-titanium oxide nanospheres sized about 5 to 10 nm. X-ray dispersion survey also suggested the high purity of the sample. In addition, the absorption rate was measured in the presence of ultrasound waves and the results indicated that the capacity of the synthesized sample to absorb Congo red is greatly dependent on the intensity power of ultrasound waves, as the absorption rate reaches 100% at powers above 30 watts.

## 1. INTRODUCTION

Shortage of water resources and the increasing expansion of industrial units has increased the production of industrial wastewater and contaminated water resources, which is considered a major socioeconomic problem. Chemicals affect water quality indicators, such as color, taste, and smell. Since these compounds may produce unwanted disinfectants such as trihalomethanes (that most of them are carcinogens), it seems necessary to control them. Industries such as textile, pulp and paper, pharmaceutical, and leather, due to the consumption of thousands of color chemicals, are among the major dischargers of color pollutants into the environment. Many of these colors are resistant to biodegradation processes. The existence of colors in the sewage of such industries prevents the sunlight from penetrating into the water and reduces the rate of photosynthetic processes in surface water [1-5]. Removal of organic compounds from aqueous environments has been widely studied [6-10]. The most common methods for removing these compounds include ozonation [11], coagulation [12], ion exchange [13], biodegradation [14], the use of membranes [15], and absorption [16]. The high cost of advanced oxidation processes and membranes, the need for a large amount of coagulants and production of high amounts of sludge in the coagulation process to remove humic acid, and low removal efficiency in degradation are some of the disadvantages of these methods. Therefore, to maintain human and environmental health, the use of inexpensive and cost-effective methods for removal of colors from the wastewater of textile and manufacturing plants is of special importance in developing countries [17]. Since the adsorption process is more inexpensive and simpler to use, it is more preferred to other methods [18]. The adsorption process is considered the most suitable method for removing color pollutants and improving the quality of industrial sewage. Because of low cost, easy design, ease of use, and insensitivity to toxic materials, it is a well-known process in wastewater reuse [19]. Recently, magnetic particles have attracted attentions as a new absorbent, which is attributed to their magnetic properties and separability by magnets, the large number of active absorption sites on their surface, and high pollutant removal efficiency [20-22]. In this regard, iron oxide nanoparticles are known as an effective absorbent for the removal of pollutants from aqueous environments and are widely used to remove organic pollutants and heavy metals from such environments [23]. Assadi *et al*. studied the removal of reactive red color from an aqueous medium using the surfactant-modified iron oxide nanoparticles. Their results showed that more than 98% of colors were removed using this method [24]. Gutierrez *et al*. conducted a review study on the removal of nickel, copper, cadmium, and chromium ions using iron oxide nanoparticles and some of researchers reported that more than 90% of ions were removed by these nanoparticles [25]. Hence, the objective of this research is to remove Congo red color from the textile industry wastewater through synthesizing $Fe_3O_4@TiO_2$ core-shell nanospheres and investigate their absorption features in the presence of ultrasound waves.

## 2. EXPERIMENTAL

## 2.1. CHEMICAL MATERIALS

Fecl$_3$.6H$_2$O, Fecl$_2$.4H$_2$O, cetyltrimethylammonium bromide (CTAB), sodium dodecyl sulfate (SDS), titanium tetra-isopropoxide (TTIP), 28% ammonium hydroxide solution, Congo red (CR), ethanol, hydrochloric acid, sodium hydroxide, and double distilled water (DDW) were the chemical used in this study. All these substances were purchased from Merck Company, Germany.

## 2.2. SYNTHESIS OF Fe$_3$O$_4$ NANOPARTICLES

At first, 2.01 gr Fecl$_3$.6H$_2$O and 0.7923 gr Fecl$_2$.4H$_2$O were added to 40 ml non-oxygenated water aerated with argon gas for 15 minutes. After a transparent solution was obtained, 45.1 ml of 0.1 molar sodium hydroxide solution was added to it. Finally, the solution pH was set to range between 11 and 12. After precipitation, 0.1 gr CTAB was added to the solution as a surfactant. The resulting mix was placed under treatment with argon and stirred at ambient temperature using a magnetic agitator. After 20 minutes, sediments were separated using a magnet and were washed firstly with water and diluted acid and then with ethanol in order to both purify them and set pH at the neutral level. The resulting sediments were dried in an oven at 130°C for 30 minutes and then were powdered to be used for the synthesis of magnetic core-shell nanoparticles.

## 2.3. SYNTHESIS OF Fe$_3$O$_4$@TiO$_2$ MAGNETICS CORE-SHELL NANOSPHERE

Similar to the stirring method, 0.7 gr CTAB and a mixture of 90 ml ethanol, 35 ml ion-free water, and 3 ml 28% ammonium hydroxide were added to 0.7 gr Fe$_3$O$_4$ prepared in the previous step. Then, the resulting mix was placed under waves at 25 watts for 20 minutes. A certain amount of TTIP was added to the mix and again placed under waves for 1 hour. The resulting sediments were separated using a magnet and then washed several times with water and ethanol in order to remove impurities. In the next step, the nanoparticles were dried at 55°C for 5 hours. A schematic view of the process of synthesizing iron oxide-titanium oxide nanospheres is shown in Figure 1.

## 3. RESULTS

Figure 2 depicts the XRD spectra of synthesized nanoparticles. The results of XRD investigation of samples confirm the inverse spinel cubic structure of Fe$_3$O$_4$. In addition, peaks of 220, 311, 400, 422, 511, and 440 are fully consistent with the standard cubic structure of magnetic iron oxide (JCPDS, No. 19-0629). It is obvious that Fe$_3$O$_4$ is the only recognizable phase and there is no other phase of iron like γ-Fe$_2$O$_3$, as one of the products of coprecipitation whose peaks are very close to this structure.

The Fe$_3$O$_4$ crystalline network consists of tetrahedron and octahedral sites in which iron ions are surrounded by 4 and 6 ions of oxygen, respectively. As it can be observed, the appeared peaks are separately similar to Fe$_3$O$_4$ structure, but the intensity of peaks was reduced due to less coverage. However, since the peaks related to TiO$_2$ crystalline structure cannot be observed, it can be concluded that titanium oxide nanoparticles amorphously cover the Fe$_3$O$_4$ surface.

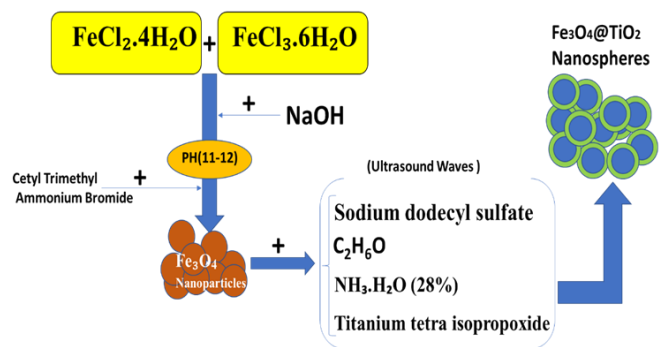

Figure 1: A schematic view of the process of synthesizing iron oxide-titanium oxide nanospheres

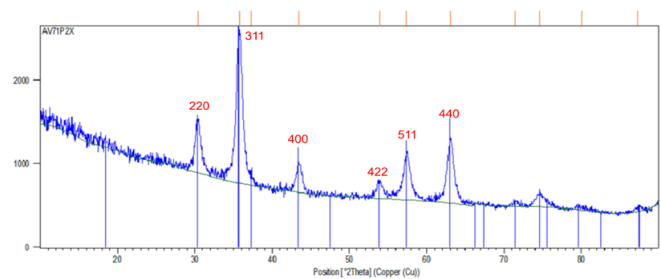

Figure 2: X-ray diffraction spectrum of iron oxide-titanium oxide core-shell nanospheres

Figure 3 shows Fe$_3$O$_4$@TiO$_2$ nanoparticles. Most particles are spherical with a mean size of 5-10 nm. It can be stated that the use of ultrasound waves in synthesis results in the production of smaller particles. This is consistent with the results of XRD.

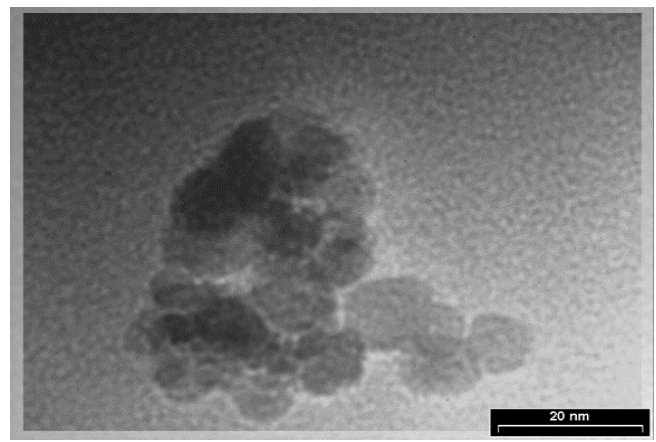

Figure 3: Images of Fe$_3$O$_4$@TiO$_2$ nanoparticles synthesized under ultrasound using TEM.

### 3.1 THE EFFECT OF SYNTHESIZED FE3O4@TIO2 NANOSPERES ON THE REMOVAL OF COLOR POLLUTANTS:

First, a certain amount of the desired absorbent at the specified pH and temperature was poured into the cell and placed under waves or stirring. To ensure the accuracy of the results, each test was repeated three times. Figure 4 depicts the chemical structure of Congo red removal in different conditions. It is noteworthy that Congo red is one of the most widely used colors in the textile industry that, because of their structure, are classified among stable and very toxic colors in terms of degradability.

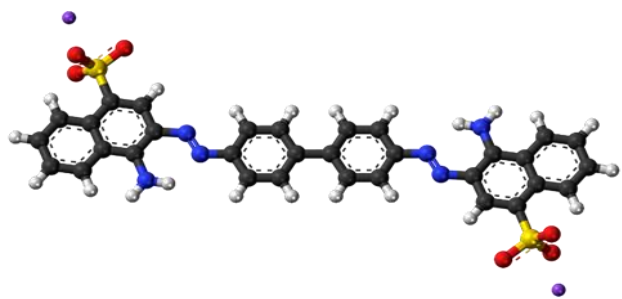

Figure 4: Chemical structure of Congo red

### 3.2. THE EFFECT OF ULTRASOUND WAVES INTENSITY:

Figure 5 shows that color absorption reduces in low intensities but increases in intensities higher than 28 watts, because the increased number of bubbles and acute conditions in the cavitation process at high intensities improve the influence on the surface and mass transfer. Therefore, the optimum intensity for studying other factors was determined to be 28 watts. The following chart has been plotted based on the power calculated by the device using calorimetry.

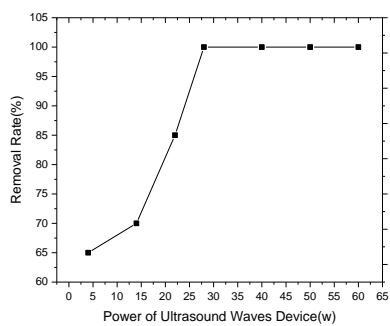

Figure 5: The effect of the device intensity on Congo red absorption (concentration: 150 mg/l, absorption rate: 0.09 gr, temperature: 21-24°C, time: 3 minutes)

In this stage, the effect of different concentrations of the absorbent (0.01 to 0.15 gr) for 65 ml of a 150 mg/l color solution was studied at a normal pH for 3 minutes. Figure 6 shows that with increasing concentration of the absorbent up to 0.15 gr, the reaction speed increases in the classical method, resulting in a color removal rate of 99.55%. In the ultrasound waves method, the reaction speed also increases with the increasing amount of the absorbent, but the color removal rate was 99.7% at the concentration of 0.09 gr.

Since the amorphous structure of titanium oxide contains $Ti^{4+}$ and $O^{2-}$ ions (based on the XRD results), superficial atoms cannot be fully coordinated and thereby the coordination unsaturation degree increases. In addition, the effect of the surfactant on the arrangement of this configuration is not negligible. The minimum configuration energy for surface atoms occurs when they are in close connection with the surroundings at the atomic scale. When the surface degree of unsaturation increases, driving forces for chemical and physical absorption will increase.

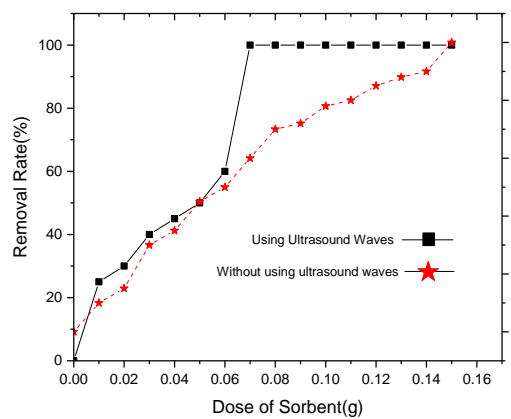

Figure 6: The effect of absorbent concentration on Congo red absorption (concentration: 150 mg/l, temperature: 21-24°C, time: 3 minutes)

the surface degree of unsaturation increases, driving forces for chemical and physical absorption will increase. Here, there are two driving forces for molecular absorption: 1- coordination of surface atoms with the surrounding atoms and 2- the interaction of the surface charge with opposite ions. It is obvious that the increased absorbent concentration leads to an increase in the absorption rate due to the increased contact surface of the absorbent, because the increased concentration of the absorbent increases the contact surface of absorbent particles. As a result, more color groups can sit on the absorbent surface and thereby the absorption rate will also increase. Moreover, the burst of cavitation bubbles causes the formation of high-pressure turbulent currents and high-speed microjets which convert larger particles into smaller ones, and this reduction in particles size increases the contact surface and also the reactivity. In addition, the disturbance caused by waves due to mechanical pressures improves the mass transfer process. The bubbles burst can also affect the infiltration and surface structure in this structure. Therefore, the optimum absorbent concentration for studying other factors was determined to be 0.1 gr.

### CONCLUSION:

The results showed that the intrinsic particle infiltration rate in the stirring method is higher than that of the ultrasound. This can be attributed to the thickening of the boundary layer in the ultrasound method due to the increased number of color molecules on the external surface of the adsorbent. Moreover, with increasing temperature due to increased ion mobility and eventually increased number of color molecules in the boundary layer and the external surface, the intrinsic particle infiltration rate reduces in the stirring method, while it increases in the ultrasound waves method because of accelerated mass transfer into the pores as a result of mechanical pressures. It can be probably stated that, in the presence or absence of waves, increased temperature causes the superiority of layer infiltration over the intrinsic particle infiltration. However, intrinsic particle infiltration is faster in the stirring method considering the thickness of the boundary layer. Nonetheless, the absorption rate in these steps is very low compared to the absorption rate on the surface. This can be attributed to interactions between color molecules and absorbent surface, that are stronger in the ultrasound waves method.

**CONFLICT OF INTEREST**

The work was partly supported by VEGA Grant No. 2/0009/16. R. Pincak would like to thank the TH division in CERN for hospitality.